\documentclass[10pt,tbtags,reqno]{amsart}

\usepackage{array}
\usepackage{graphicx}
\usepackage{tabmac}
\usepackage{mathrsfs}
\usepackage{latexsym}
\usepackage{amsthm,amsopn,amsfonts,amssymb,epsfig,stmaryrd,color}
\usepackage[active]{srcltx}

\makeatletter
\renewcommand{\subsubsection}{\@startsection
{subsubsection}
{3}
{0mm}
{\baselineskip}
{-0.5\baselineskip}
{\normalfont\normalsize\bfseries}}
\makeatother

\newlength{\eqboxstorage}

\makeatletter
\renewcommand{\subsubsection}{\@startsection
{subsubsection}
{3}
{0mm}
{\baselineskip}
{-0.5\baselineskip}
{\normalfont\normalsize\bfseries}}
\makeatother

\theoremstyle{remark}

\def\sap{\bigskip}

\def\la{{\lambda}}

\def\cal L{{\mathcal L}}

\def\N{{\mathbb N}}
\def\Z{{\mathbb Z}}
\def\sap{\bigskip}

\def\b\beta
\def\aa{\alpha}

\newcommand{\tcercle}[1]{\ensuremath{\setlength{\unitlength}{1ex}\begin{picture}(2.8,2.8)\put(1.4,1.4){\circle{2.7}\makebox(-5.6,0){#1}}\end{picture}}}

\let\d\partial

\def\R{{\rangle}}

\def\G{{\mathcal G}}

\def\B{{\mathcal B}}
\def\F{{\mathcal F}}

\let\la\lambda
\let\La\Lambda
\let\Om\Omega
\let\om\omega
\let\ta\theta

\let\Rw\Rightarrow

\let\ti\tilde
\newcommand{\LL}{\ensuremath{\langle\!\langle}}
\newcommand{\RR}{\ensuremath{\rangle\!\rangle}}

\def\cd{{\circledast}}

\def\cb#1{{\color{blue}#1}}
\def\co#1{{\color{red}#1}}
\def\cov#1{{\color{green}#1}}

\let\l\left
\let\r\right
\let\e\epsilon

\let\a\alpha

\def\lev{{\rm level}}
\def\beq{\begin{equation}}
\def\eeq{\end{equation}}
\def\bea{\begin{align}}
\def\eea{\end{align}}

\begin{document}


\title{Super-Whittaker vector at $c=3/2$}

\author{Patrick Desrosiers}
\address{Instituto de Matem\'atica y F\'{\i}sica, Universidad de
Talca, 2 norte 685, Talca, Chile.}
\email{desrosiers@inst-mat.utalca.cl}

\author{Luc Lapointe}
\address{Instituto de Matem\'atica y F\'{\i}sica, Universidad de
Talca, 2 norte 685, Talca, Chile.}
\email{lapointe@inst-mat.utalca.cl}

\author{Pierre Mathieu}
\address{D\'epartement de physique, de g\'enie physique et
d'optique, Universit\'e Laval,  Qu\'ebec, Canada,  G1V 0A6.}
\email{pmathieu@phy.ulaval.ca}


 \begin{abstract}
The degenerate Whittaker vector of the superconformal algebra can be represented
  in terms of  Jack superpolynomials. However, in this representation the norm of the Whittaker vector  involves a scalar product with respect to which the Jack superpolynomials are not orthogonal. In this note, we point out that this defect can be cured at $c=3/2$ by means of a trick specific to the supersymmetric case. 
At $c=3/2$, we  thus end up with a closed-form expression for the norm of the  degenerate super-Whittaker vector. Granting the super-version of the AGT conjecture, 
this closed-form expression should be equal to the $\Z_2$-symmetric  $SU(2)$ pure-gauge instanton partition function -- the corresponding
equality taking the form of a rather nontrivial combinatorial identity. 
\end{abstract}

\maketitle




 \maketitle

\section{Introduction}

In the context of the AGT conjecture \cite{AGT} applied to asymptotically free theories and in the absence of matter, Gaiotto \cite{Gai} has considered  vectors not annihilated by $L_1$ -- dubbed degenerate  Whittaker vectors. An elegant representation for these  vectors has been obtained as a sum of Jack polynomials \cite{Yan} {(see also \cite{AY})}. However, the immediate usefulness of the resulting expression is not clear since the norm of the Whittaker vector -- the degenerate conformal block  \cite{MMM} --  is then expressed in terms of a scalar product with respect to which the Jack polynomials $P^{(\alpha)}_\la$ are not orthogonal, namely
\beq \LL\, P^{(\alpha)}_\la \,| \,P^{(\alpha)}_{\om} \,\RR_{-2\alpha}\,  ,\eeq
whose expression is  not known. 

These results have been generalized to the superconformal case in \cite{DLMsvir}. There, it is shown that the degenerate super-Whittaker vector is represented as a sum of Jack superpolynomials (sJacks)  $P^{(\alpha)}_\La$ (where $\La$ is now a 
superpartition instead of a partition) and that its norm is formulated in terms of the scalar product 
\beq \LL\, P^{(\alpha)}_\La \,| \,P^{(\alpha)}_{\Om} \,\RR_{-\alpha}\,  ,\eeq
whose explicit form is again unknown. However, notice that the value $-2\a$ in the Virasoro case gets replaced by $-\a$ in the superconformal case. This numerical simplification allows us to derive a closed-form expression for the norm of 
super-Whittaker vector at $\a=1$, that is, at the value $c=3/2$ of the central charge. In order to place this result in context, we first briefly review 
certain relevant results of \cite{DLMsvir}.

\section{Superpolynomial representation of the degenerate  NS Whittaker vector}

\subsection{The degenerate  NS Whittaker vector} 
In the Neveu-Schwarz (NS) sector, the superconformal degenerate Whittaker vector  in the Verma module with  highest weight state $|h\R^\mathrm{NS}$, is defined order by order from the following recursion relations \cite{BF}: 
\beq \label{whittRSa}    G_{\frac{1}{2}}|h\R_k^\mathrm{NS}= |h\R_{k-\frac{1}{2}}^\mathrm{NS},\qquad G_{\frac{3}{2}}  |h\R_k^\mathrm{NS}=0,  \qquad \forall\, k\in \frac{\mathbb{N}}{2}.
\eeq
where $|h\R^\mathrm{NS}_{k}$ denotes a descendant of {$|h\R^\mathrm{NS}\equiv |h\R_0^\mathrm{NS}$} at level $k$.

Through the free-field representation of the superconformal algebra and the symmetric polynomial
representation of the free modes (see \cite{AMOSa,SSAFR,CJ} in the Virasoro case), we can represent the  NS Whittaker vector
in terms of superpolynomials. In the NS-sector,
the free-field representation of the superconformal algebra
\begin{align}\label{ffLG}
L_n&=-\gamma (n+1)a_n+\frac{1}{2}\sum_{m\in\mathbb{Z}}:a_ma_{n-m}:+\frac{1}{4}\sum_{k\in\mathbb{Z}+\frac{1}{2}}(n-2k):b_kb_{n-k}: \nonumber
\\
G_k&=-2\gamma \left(k+\frac{1}{2}\right)b_k+\sum_{m\in\mathbb{Z}}a_mb_{k-m},
\end{align}
where
\beq\label{cacao}
[a_n,a_m]=n\delta_{n+m,0}\qquad \text{and} 
\qquad \{b_k,b_l\}=\delta_{k+l,0}.
\eeq
The indices $n,m$ are integer, while $k$ and $l$ are half-integer. The  central charge is  $c=3/2-12\gamma^2$,
which we parametrize as
\beq
c=\frac{15}2-3\left(\a+\frac1{\a}\right) \qquad \Longrightarrow \qquad \gamma=\frac12\left(\sqrt{\a}-\frac1{\sqrt{\a}}\right).
\eeq

The highest-weight states in the Fock space $\mathscr{F}$
 are characterized by a complex number $\eta$ and satisfy
 \beq
 a_0|\eta\R=\eta|\eta\R,\qquad \text{and}\quad a_n|\eta\R=b_k|\eta\R=0, \; \quad \forall n,k>0.\eeq
The full Fock space  $\mathscr{F}$   is the linear span over $\mathbb{C}$ of all monomials 
\beq b_{-k_1}\cdots b_{-k_p}a_{-n_1}\cdots a_{-n_q}|\eta\R ,\qquad k_i, l_i>0 \, . \eeq
Finally, the conformal dimension of $|\eta\rangle$ is 
$h=\tfrac12{\eta^2}-\gamma\eta$.
 
 \subsection{Fock states and symmetric superpolynomials}

The correspondence between Fock states and symmetric functions is made via the generalization to superspace of the
power-sum symmetric functions.  The $n$-th power-sum in infinitely many variables, denoted $p_n$, and its fermionic partner $\tilde p_n$ are defined as \cite{DLM6}
\beq
  p_n=\sum_ix_i^n \quad(n>0)\qquad\text{and}\qquad\tilde p_n=\sum_i\ta_i x_i^{n},\quad (n\geq0),
\eeq
where $\ta_1,\theta_2,\ldots$ are anticommuting variables. Both $p_n$ and $\tilde p_n$ are symmetric superpolynomials, that is, polynomials in the variables $x_1, x_2,\ldots$ and $\theta_1, \theta_2,\ldots$ that remain unchanged under any simultaneous permutation of the form $(x_i,\theta_i)\leftrightarrow (x_j,\theta_j)$.  Any element of the space 
 $\mathscr{R}$ of all symmetric superpolynomials 
in infinitely many variables with coefficients in $\mathbb C$ can be uniquely written as a polynomial in $p_1,p_2,\ldots$ and $\tilde p_0, \tilde p_1,\ldots$ (see for instance  \cite{DLM6}).

The announced correspondence between the free-field modes and the differential operators acting on $\mathscr{R}$ reads:
\begin{align}\label{scor}
&a_{-n}\longleftrightarrow\frac{(-1)^{n-1}}{\sqrt{\alpha}}\, p_n 
&&a_n\longleftrightarrow  n(-1)^{n-1}\sqrt{\alpha} \,\frac{\partial}{\partial{p_n}}
\nonumber\\
&b_{-k}\longleftrightarrow \frac{(-1)^{k-1/2}}{\sqrt{\alpha}}\, \tilde p_{k-1/2}
&&b_k \longleftrightarrow (-1)^{k-1/2}\sqrt{\alpha}\, \frac{\partial}{\partial {\tilde p_{k-1/2}}}, 
\end{align}
where $k,n>0$ and $\a$ is a non-zero free parameter.
This implies the correspondence
\begin{align} \label{cors}
b_{-k_1}\cdots b_{-k_m} a_{-n_1}\cdots a_{-n_p}\, |\eta\R& \longleftrightarrow\zeta\, \tilde p_{k_1-\frac12}\cdots \tilde p_{k_m-\frac12}\, p_{n_1}\cdots p_{n_p},
\end{align}
with
 $k_i> k_{i+i}\geq \tfrac12 $ and $ n_i\geq n_{i+1}\geq 1$,
and $\zeta$ is a constant  read off \eqref{scor}.
We then relabel the indices as
\begin{align} \label{indi}
 k_i-\tfrac12&= \La_i \quad  \quad\, \text{for}\quad 1\leq i\leq m,&& (\Rw\; \La_{i}>\La_{i+1}\geq 0 \quad\text{for}\quad1\leq i\leq m-1),\nonumber\\
 n_i&=\La_{i+m}\quad \text{for}\quad  1\leq i\leq p=\ell-m,&& (\Rw\; \La_{i}\geq\La_{i+1}\geq 1 \quad \text{for}\quad m+1\leq i\leq \ell-1).
 \end{align}

Thus, any state in $\mathscr{F}$ is in correspondence with a polynomial in $\mathscr{R}$  indexed by two partitions: $\La^a=(\La_1,\ldots,\La_m)$, whose elements are strictly decreasing   with $\La_m\geq 0$, and  $\La^s=(\La_{m+1},\ldots,\La_\ell)$, which is a standard partition with $\ell$ non-zero elements.  Together, the partitions $\La^a$ and $\La^s$ form the superpartition $\La=(\La_{1},\ldots,\La_m;\La_{m+1},\ldots,\La_\ell)$ (see \cite{DLM6} and references therein). The non-negative integer  $m$ is called the fermionic degree of the superpartition $\La$   while its bosonic degree is given by $|\La|=\sum_{i=1}^\ell\La_i$. 
From $\La$, we construct two partitions: $\La^*$, which is obtained by removing the semi-coma and reordering the parts in non-increasing order, and $\La^\cd$, which is obtained similarly but from $(\La_{1}+1,\ldots,\La_m+1;\La_{m+1},\ldots,\La_\ell)$. Clearly, $\La$ is uniquely specified by the pair $(\La^*,\La^\cd)$. The diagram of $\La$ is that of $\La^*$ but with circles added to the end of the $m$ rows for which $\La^\cd_i-\La^*_i=1$.  The conjugate of $\La$, denoted $\La'$, is the superpatition associated with the transposed diagram of $\La$ obtained by reflecting along the main diagonal.    For instance, for $\La
=(3,1,0;2,1)$, we have
\beq \label{exa}
\La
: \quad {\tableau[scY]{&&&\bl\tcercle{}\\&\\&\bl\tcercle{}\\ \\
    \bl\tcercle{}}}\qquad
     \La^\cd:\quad{\tableau[scY]{&&&\\&\\&\\\\ \\ }} \qquad
         \La^*:\quad{\tableau[scY]{&&\\&\\ \\ \\ }}   \qquad \La'
: \quad {\tableau[scY]{&&&&\bl\tcercle{}\\&&\bl\tcercle{}\\ \\\bl\tcercle{}}}\;. \eeq

\subsection{sJack representation of the degenerate  NS Whittaker vector}
Bases of the space $\mathscr{R}$ can be naturally indexed by superpartitions \cite{DLM6,DLM7,DLMeva}.  For instance, the power-sum basis is defined as
   \beq 
\label{pdef}
p_\La=\tilde p_{\La_1}\cdots \tilde p_{\La_m}p_{\La_{m+1}}\cdots p_{\La_{\ell}} .\eeq
Thus, any symmetric superpolynomial can be expanded as a sum over the $p_\La$'s, and in particular, the Jack superpolynomials (sJack) $P_\La^{(\a)}$.

The  degenerate NS super Whittaker vector at level $k$ can thus be represented as a sum over sJacks as
\beq \label{eqwihittskacks}  |h\R^\mathrm{NS}_{k} \longleftrightarrow W_k=\sum_{\La,\,\lev ( \La)=k}w_\La P_\La ,
\eeq 
where  the level of the superpartition $\La$ is defined 
as
\beq \label{deflev}
 \lev(\La)=\frac12(\,{|\Lambda^*|+|\La^\circledast|}\,)=|\La^*|+\frac{m}2.\eeq
 For instance, $(3,1,0;2,1)$ has level $17/2$.

The free field representation \eqref{ffLG} and the correspondence \eqref{scor} immediately imply that the generators $G_k$ and $L_n$  can be represented as differential operators acting on the space $\mathscr{R}$ of symmetric  superpolynomials. Let us denote these differential representations by $\mathcal{G}_r$ and $\mathcal{L}_n$ respectively. For instance, we have
\begin{align}
&\mathcal{G}_{\frac12}= (\bar\eta-2\bar\gamma) \frac{\d}{\d \ti p_0} +\sum_{n>0}(n\,\ti p_{n-1}\,  \frac{\d}{\d  p_n} - p_n\,  \frac{\d}{\d \ti p_n})\label{g12} \\
\G_{\frac32}&=-(\bar\eta-4\bar\gamma) \frac{\d}{\d \ti p_1}+\a \frac{\d}{\d  p_1} \frac{\d}{\d \ti p_0}-\sum_{n\geq 2}n\,\ti p_{n-2}\, \frac{\d}{\d p_n}+\sum_{n>0}p_n\, \frac{\!\!\!\!\d}{\d \ti p_{n+1}},
\end{align}
where $\bar\eta =\sqrt{\alpha}\eta$ and $\bar\gamma =\sqrt{\alpha}\gamma$.

   The coefficients $w_\La=w_\La(\a,\bar\eta)$ in \eqref{eqwihittskacks} are fixed by the initial condition $W_0=1$ and the relations 
\beq \mathcal{G}_{\frac{1}{2}}W_k= W_{k-\frac{1}{2}}\, ,\qquad \mathcal{G}_{\frac{3}{2}}W_k=0,  \qquad \forall \, k>0\, . \eeq 
The expression for the coefficients in \eqref{eqwihittskacks} has been reported in \cite{DLMsvir}: 
\beq \label{coeffw}
w_\La(\a,\bar\eta)=\frac{(-1)^{\binom{m}{2}} \a^{|\La^*|}}{ \bar \eta \, h^\uparrow_\La}
\left[\prod_{\substack{(i,j)\in \La^\cd \\ (i,j) \neq (1,1)}}\frac1{\bar\eta+i-\a j}\right]\, A(\La), \eeq
where (using $\lambda=\Lambda^*$)
\beq\label{defA}
A(\La)= \prod_{(i,j)\in \La^*_{\text{nr}}}
\frac{\bigl[2\bar\eta+1+i+\la_j'-\a(1+j+\la_i)\bigr]}{2\bar\eta+1+2i-\a(1+2j)}
\prod_{(i,j) \in \mathcal{F}\La} \frac{1}{2\bar\eta+1+i+\la_j'-\a(1+j+\la_i)}.
 \eeq
In the previous equation, $\La^*_{\text{nr}}$ stands for the cells of $\La^*$ that are not removable corners 
(a removable corner of a partition is a cell that lies both at the end of its row and at the end of its column),
and $\mathcal{F}\La$ denotes the set of boxes $s=(i,j)$ in the diagram of $\La$ that belong at the same time in a fermionic row and in a fermionic column (a row/column is said to be fermionic if it terminates with a circle).
 For example, in the  diagram of $\La=(3,1,0;2,1)$ displayed in \eqref{exa}, the boxes in  $\F\La$ are $(1,1),\,(1,2)$ and $(3,1)$.
  Finally, $h^\uparrow_\La$ is defined below in \eqref{defhook}.

Albeit conjectural, the closed-form expression \eqref{coeffw}--\eqref{defA} of the coefficients $w_\La$
 has been checked extensively via its norm (cf. \eqref{DLMconj} below).

\subsection{Norm of the degenerate  NS Whittaker vector}
The motivation for studying 
the degenerate Whittaker vector is rooted  in the observation 
that the square of its norm, $\big( \,|k \rangle \, , \, |k \rangle \, \big)_{c,h}$, is equal to the degenerate limit of the four-point conformal block at level $k$ \cite{Gai,MMM,BF}. 
Here $|k\rangle\equiv|h \rangle_{k}^\mathrm{NS}$ denotes the level-$k$ degenerate Whittaker vector,  and  $( \,  \,,\, )_{c,h}$ stands for the usual Hermitian Shapovalov form  on the NS highest-weight module.  The latter form is non-degenerate
and characterized, up to a constant, by 
the following invariance property:
\beq  \label{invform} \big( \, G_r|\La \rangle \, , \, |{\Om} \rangle \, \big)_{c,h} = \big( \,|\La \rangle \, , \, G_{-r}|{\Om} \rangle \, \big)_{c,h}\eeq
for all half-integers $r$ and for all basic states $|\La \rangle,{|\Om \rangle}$ of the NS highest-weight module, where
we made use of the shorthand notation
\beq\label{etat}
|\La\R=G_{-\La_1-\frac12}\cdots G_{-\La_m-\frac12} L_{-\La_{m+1}}\cdots L_{-\La_\ell}\, |h\R^\mathrm{{NS}}.
\eeq

Our interest is to rephrase the norm squared $\big( \,|k \rangle \, , \, |k \rangle \, \big)_{c,h}$ in the language of symmetric superpolynomials.  
Given eqs. \eqref{ffLG} and \eqref{scor}, we already know how to represent any element of the highest-weight module over the NS sector as an element of the space $\mathscr{R}$ of symmetric superpolynomials:
\beq G_{-r_1}\cdots G_{-r_m}L_{-n_1}\cdots L_{-n_p}|h\R^\mathrm{{NS}}\longleftrightarrow \mathcal{G}_{-r_1}\cdots \mathcal{G}_{-r_m}\mathcal{L}_{-n_1}\cdots \mathcal{L}_{-n_p}(1)=\sum_{\La} u_\La P_\La^{(\alpha)},\eeq
 where the sum runs over all  superpartitions whose  level is equal to $\sum_i r_i+\sum_j n_j$ and where the $u_\La$'s {denote} complex {coefficients} depending upon $\alpha$ and $\eta$.

The space $\mathscr{R}$ is naturally equipped with the  scalar product $\LL \, \,| \,\, \RR_\alpha$ defined as
\begin{equation} \label{scap} \LL \, 
{p_\La} \, | \, {p_\Om }\, \RR_\alpha=(-1)^{\binom{m}2}\, \alpha^{{\ell}(\La)}\, z_{\La^s}
\delta_{\La,\Om}\,,\qquad \text{where}\qquad
z_{\La^s}=\prod_{i \geq 1} i^{n_{\La^s}(i)} {n_{\La^s}(i)!}\, ,
\end{equation}
with $n_{\La^s}(i)$ the number of parts in $\La^s$ equal to $i$.  It turns out that the sJacks are orthogonal with respect to the scalar product \eqref{scap}:
\begin{equation} \label{ortho} \LL \, 
P_\La^{(\a)} \, | \,P_\Om^{(\a)}\, \RR_\alpha=0 \quad\text{when}\quad \La\ne\Om\, .
\end{equation}
In order to make  this scalar product   compatible with the invariance property \eqref{invform}, we make a slight change of parametrization, 
\beq \label{para}
\eta=\rho+\gamma \qquad \Longrightarrow \qquad h=\frac{1}{2}(\rho+\gamma)(\rho-\gamma),\eeq	
and set
$\varphi(\rho)=- \rho$ and $\varphi(\gamma)=\gamma$, 
where $\varphi$ denotes complex conjugation.
We then define 
\beq  \LL \, f\,| \,g\, \RR^\varphi_\beta\equiv \LL \,\varphi(f) \,| \,g\, \RR_\beta, \eeq
which is a Hermitian form on $\mathscr{R}$, where $\beta$ is a function of $\alpha$ which
 is determined by enforcing  the invariance
\eqref{invform}. A close examination reveals that  $\beta=-\alpha$ \cite{DLMsvir}.
Therefore, 
 if $|\La\R \longleftrightarrow f $ and $|\Om\R \longleftrightarrow g$, then
\beq \label{equivforms}\big( \,|\La \rangle \, , \, |\Om \rangle \, \big)_{c,h} =  \,
\LL\, f \,| \, g \,\RR_{-\alpha}^{\varphi} \, .\eeq
By writing 
\beq W_k(\alpha,\bar \rho)=\sum_{\La, \mathrm{level}(\La)=k} w_\La(\alpha,\bar \rho)P^{(\alpha)}_\La\,,\eeq
where we now use $w_\La(\alpha,\bar \rho)$  instead of $w_\La(\alpha,\bar \eta)$ as in \eqref{coeffw}
to emphasize that $w_\La$ only depends on $\alpha$ and $\bar \rho$ (given that 
$\bar \eta=\bar \rho+\bar \gamma=\bar \rho+ (\alpha-1)/2)$), we have
\beq \big( \,|k \rangle \, , \, |k \rangle \, \big)_{c,h}= \,{\LL} \, W_k(\alpha,\bar \rho)\,\big| \, W_k(\alpha,\bar \rho)\, {\RR}^\varphi_{-\alpha}= \,{\LL} \, W_k(\alpha,-\bar \rho)\,\big| \, W_k(\alpha,\bar \rho)\, {\RR}_{-\alpha}\, ,\eeq
so that,\footnote{Our expressions for the left-hand side differ slightly from those of \cite{BF}, listed there for $\tfrac12\leq k\leq \tfrac52$ (and which we label BF):
 $$\big( \,|k \rangle \, , \, |k \rangle \, \big)^{\text{BF}}_{c,h} =\big( \,|k \rangle \, , \, |k \rangle \, \big)_{c,h} \times 4^{-\lfloor k \rfloor}.$$}
\beq\label{DLMconj} \big( \,|k \rangle \, , \, |k \rangle \, \big)_{c,h} 
=\,{ \sum_{\substack{\La,\Om\\\textrm{level}(\La)=\lev(\Om)=k}}  }
w_\La(\alpha,-\bar \rho)\, w_{\Om}(\alpha,\bar \rho)\,\LL\, P^{(\alpha)}_\La \,| \,P^{(\alpha)}_{\Om} \,\RR_{-\alpha}\,  .
\eeq
This equality has been tested up to  level 13/2. 
Note that the norm on the left-hand side is equal to the coefficient, in $|k\R$, of the term containing solely the operators $G_{-\frac12}$ and $L_{-1}$. More explicitly, if
\beq |k\R=\sum_{\substack{\La\\   \lev(\La)=k}}c_\La|\La\R,\eeq
where we used the notation \eqref{etat}, then
\beq \big( \,|k \rangle \, , \, |k \rangle \, \big)_{c,h} =c_{\La^0}\qquad \text{where}\qquad |\La^0\R=G_{-\frac12}^{2(k-\lfloor k\rfloor)}\, L_{-1}^{\lfloor k \rfloor}
|h\R^\mathrm{{NS}}.\eeq
This is an easy consequence of the fact that the only non-vanishing actions of positive super-Virasoro modes on $|k\R$ are $G_{\frac12}$ and $L_1$. (This is analogous to the situation in the Virasoro case \cite{MMM,BF}.)

 Given that $\LL\, P^{(\alpha)}_\La \,| \,P^{(\alpha)}_{\Om} \,\RR_{-\alpha}$ is not known, the usefulness of expression \eqref{DLMconj} is questionable. However, we will see that at $\a=1$ it can be turned into a closed-form expression.

\section{Norm of the NS Whittaker vector at $c=3/2$}

\subsection{Duality transformations}
By replacing $\alpha$ by $-\alpha$ in the scalar product \eqref{scap}, we get
\beq
\LL  p_{\Lambda}\,| \, p_{\Omega} \RR_{-\alpha} = 
(-1)^{\binom{m}{2}+m+\ell(\Lambda^s)}  \alpha^{\ell(\Lambda)}z_{\Lambda^s}\, \delta_{\Lambda \Omega} 
\eeq
where we used the relation $\ell(\Lambda)=m+\ell(\Lambda^s)$.
Introduce the operator $\hat \omega$ whose action on the elementary power-sums is defined as \cite{DLM7}
\beq
\hat \omega (p_r) = (-1)^{r-1} p_r \qquad {\rm and} \qquad \hat \omega (\tilde p_r) = (-1)^{r} \tilde p_r,
\eeq
so that on the full power-sum, we get
\beq
\hat \omega p_{\Lambda} = (-1)^{|\Lambda^*|+\ell(\Lambda^s)} p_{\Lambda} .
\eeq
We can thus relate the scalar product at $-\a$ to the one evaluated at $\a$ at the price of acting with $\hat\om$ on one of the terms:
\beq
\LL  p_{\Lambda}\,| \, p_{\Omega} \RR_{-\alpha} = (-1)^{|\Lambda^\circledast|}
 \LL  \hat \omega p_{\Lambda}\,| \, p_{\Omega} \RR_{\alpha} .
 \eeq
 
Since the sJacks can be expanded linearly in terms of the power-sums, this readily implies
\beq
\LL  P_{\Lambda}^{(\alpha)}\,| \, P_{\Omega}^{(\alpha)} \RR_{-\alpha} = (-1)^{|\Lambda^\circledast|}
 \LL  \hat \omega P_{\Lambda}^{(\alpha)}\,| \, P_{\Omega}^{(\alpha)} \RR_{\alpha} .
\eeq
This is still not  a convenient expression since there is no known
explicit expansion of $\hat \omega P_{\Lambda}^{(\alpha)}$ in terms of sJacks.
We thus consider a further simplification. Notice that 
\beq\label{ppwa}
 \LL  \hat \omega p_{\Lambda}\,| \, p_{\Omega} \RR_{\alpha} = \LL  \hat \omega_\a  p_{\Lambda}\,| \, p_{\Omega} \RR_{\alpha=1} \, ,
 \eeq
 where
 \beq
\hat \omega_\a (p_r) = (-1)^{r-1}\a p_r \qquad {\rm and} \qquad \hat \omega_\a (\tilde p_r) = (-1)^{r} \a\, \tilde p_r.
\eeq
This again implies that
\beq
\LL  \hat \omega P_{\Lambda}^{(\alpha)}\,| \, P_{\Omega}^{(\alpha)} \RR_{\alpha} = 
 \LL  \hat \omega_\alpha P_{\Lambda}^{(\alpha)}\,| \, P_{\Omega}^{(\alpha)} \RR_{\alpha=1} .
 \eeq
Explicitly, the action of $\hat\om_\a$ on $P_{\Lambda}^{(\alpha)}$ reads \cite{DLM7}:
 \beq
 \hat \omega_\alpha P_{\Lambda}^{(\alpha)}
= (-1)^{\binom{m}{2}} j_{\Lambda}(\alpha)    P_{\Lambda'}^{(1/\alpha)}, 
\eeq
where 
\beq\label{normP}
\LL P_{\Lambda}^{(\alpha)}\,| \, P_{\Lambda}^{(\alpha)}\RR_\a=(-1)^{\binom{m}{2}} j_\La(\a).\eeq
This normalization factor takes the form \cite{LLN,DLMeva}
\beq  j_\La(\a)=\a^m\frac{h^\uparrow_\La}{h^\downarrow_\La},\label{norm}
\eeq 
 where
 $h^{\uparrow\downarrow}_\La$ 
 are defined as follows: 
\begin{align}\label{defhook}
&h^\uparrow_\La=\prod_{s\in\B\La} h^\uparrow_\La(s),\qquad h^\uparrow_\La(s)=l_{\La^\cd}(s)+\a(a_{\La^*}(s)+1),\nonumber\\
&h^\downarrow_\La=\prod_{s\in\B\La} h^\downarrow_\La(s),\qquad h^\downarrow_\La(s)=l_{\La^*}(s)+1+\a\,a_{\La^\cd}(s).
\end{align}
In the previous expressions,
$\mathcal{B}\La$ corresponds to the boxes $s=(i,j)$ in the diagram of $\La$ that do not belong to $\mathcal{F}\La$ \cite{DLMeva}.
Given the box $s=(i,j)$
($i$-th row and $j$-th column) of a partition $\lambda$, 
the quantities $a_\la(s)$ and $l_\la(s)$ are defined as 
\beq a_\la(s)=\la_i-j\qquad\text{and}\qquad l_\la(s)=\la_j'-i,\eeq
where $\la'$ stands for the conjugate of $\la$, obtained by interchanging rows and columns.

Summing up, we have obtained the sequence of equalities:
\begin{align}\label{enbref}
\LL  P_{\Lambda}^{(\alpha)}\,| \, P_{\Omega}^{(\alpha)} \RR_{-\alpha} &= (-1)^{|\Lambda^\circledast|}\LL  \hat \omega P_{\Lambda}^{(\alpha)}\,| \, P_{\Omega}^{(\alpha)} \RR_{\alpha} \nonumber\\ &= (-1)^{|\Lambda^\circledast|}
 \LL  \hat \omega_\alpha P_{\Lambda}^{(\alpha)}\,| \, P_{\Omega}^{(\alpha)} \RR_{\alpha=1} \nonumber\\ &
= (-1)^{|\Lambda^\circledast|+\binom{m}{2}} j_{\Lambda}(\alpha)  \LL   P_{\Lambda'}^{(1/\alpha)}\,| \, P_{\Omega}^{(\alpha)} \RR_{\alpha=1} .
\end{align}
In other words, by introducing the operation $\hat \om$, the scalar product 
evaluated at $-\a$ is transformed into the one evaluated at $\a$, and by trading $\hat\om$ for $\hat\om_\a$, this scalar product is then changed into the one evaluated at $\a=1$.
Therefore, if we had an expression for $P_{\Omega}^{(\alpha)}$ (which would obviously provide one for $P_{\La'}^{(1/\alpha)}$) in terms of the Schur analogs $P_{\Omega}^{(1)}$, or even in terms of the power-sums $p_\Om$, we could obtain the norm of the degenerate super-Whittaker vector in closed-form for any value of $\a$.  This is an  interesting combinatorial problem which deserves further study.
However, our immediate purpose is to point out a dramatic simplification that occurs when  $\alpha=1$, that is, at $c=3/2$. 

\subsection{The case $\a=1$}
At $\a=1$, the orthogonality condition together with the normalization \eqref{normP} yield
\beq
  \LL   P_{\Lambda'}^{(1)}\,| \, P_{\Omega}^{(1)} \RR_{\alpha=1} =(-1)^{\binom{m}{2}} j_{\La'}(1) \, \delta_{\Lambda' \Omega},
\eeq
so that the last equality in \eqref{enbref} reduces to
\beq
\LL  P_{\Lambda}^{(1)}\,| \, P_{\Omega}^{(1)} \RR_{-1} = (-1)^{|\Lambda^\circledast|}j_{\La}(1)j_{\La'}(1) \, \delta_{\Lambda' \Omega}=(-1)^{|\Lambda^\circledast|} \delta_{\Lambda' \Omega} ,
\eeq
since it is easily checked that the product $j_{\La}(1)j_{\La'}(1)$ reduces to 1.

When $\a=1$, we have $\gamma=0$ and $\bar \eta =\bar \rho=\rho$.  Hence
\eqref{DLMconj} can be written in this case as
\begin{align}
\label{DLMconj32}
\big( \,|k \rangle \, , \, |k \rangle \, \big)_{c=3/2,h} 
&=\,{ \sum_{\substack{\La,\Om\\\textrm{level}(\La)=\lev(\Om)=k}}  }
w_\La(1,-\rho)\, w_{\Om}(1, \rho)\,(-1)^{|\Lambda^\circledast|} \delta_{\Lambda' \Omega}  \nonumber\\
&=\,{ \sum_{\substack{\La\\\textrm{level}(\La)=k}}  }
w_\La(1,-\rho)\, w_{\La'}(1, \rho)\,(-1)^{|\Lambda^\circledast|}.
\end{align}
In the previous expression, the product $w_\La(1,-\rho)\, w_{\La'}(1, \rho)$ 
can be written compactly in the form 
\begin{align}
w_\La(1,-\rho)\, w_{\La'}(1, \rho)\,
&=
 \frac{(-1)^{|\Lambda^\circledast|+\binom{m}{2}}}
 {4^{|\La^*_{\text{nr}}|}\left[h^\uparrow_\La h^\uparrow_{\La'}\right]_{\a=1}}\,
\prod_{{(i,j)\in \La^\cd}}\frac{1}{(\rho-i+j)^{2+2\e^r_{ij}}}
\frac{ \prod_{(i,j)\in \La^*_{\text{nr}}}(2\rho-i-\la_j'+j+\la_i)^2}
{\prod_{(i,j) \in \mathcal{F}\La}(2\rho-i-\la_j'+j+\la_i)^2}
\end{align}
where 
in the last term $\lambda=\Lambda^*$, and 
where $\e^r_{ij}=1$ if $(i,j)\in \La^*_{\text{nr}}$ and 0 otherwise.

\subsection{A digressing remark}
The exact expression for the super-Whittaker vector depends crucially upon the proper choice of the precise proportionality coefficient relating the free-field modes $a_{-n}$ to $p_n$ and $b_{-r}$ to $\ti p_{r-1/2}$.
 For instance, in the Virasoro case, the required relationship is (with $n>0$)
\beq\label{choix}
a_{-n}\longleftrightarrow\frac{(-1)^{n-1}}{\sqrt{2\alpha}}\, p_n\qquad\left(\text{so that}\qquad 
a_n\longleftrightarrow  n(-1)^{n-1}\sqrt{2\alpha} \,\frac{\partial}{\partial{p_n}}\right)
\eeq
which differs  from the corresponding relationship in the supersymmetric version \eqref{scor} in that $\sqrt{\a}$ is replaced by $\sqrt{2\a}$. To be clear, if we let
\beq\label{defo}
a_{-n}\longleftrightarrow\frac{(-1)^{n-1}}{{\kappa}\sqrt{\alpha}}\, p_n\qquad \text{and}\qquad
a_n\longleftrightarrow  (-1)^{n-1}n {\kappa}\sqrt{\alpha} \,\frac{\partial}{\partial{p_n}}
\eeq
we obtain an expansion in terms of Jack polynomials but in general the expansion coefficients are very complicated and do not factorize. This factorization is observed only for {$\kappa=\sqrt{2}$ 
, which is equivalent to the choice of parametrization made in \eqref{choix} \cite{MY}}.\footnote{If we replace $\kappa$ in \eqref{defo} by a parameter  $\kappa_n$ depending  on $n$, then we  observe the factorization  not only for $\kappa_n=\sqrt{2}$, but also for $\kappa_n =(-1)^{n-1}\sqrt{2}/\alpha$.   The correspondence associated with the latter value is  $
a_{-n}\longleftrightarrow {\sqrt{\alpha/2}}\, p_n$ and $a_n\longleftrightarrow  n\sqrt{\alpha/2} \,\frac{\partial}{\partial{p_n}}$, which {is that} used in \cite{AMOSa,Yan}.     However, we easily get this second correspondence {from the first one} by acting with $\hat \om_\alpha$ on the {symmetric-function-side} of \eqref{choix}.  Moreover, $\hat \om_\alpha (\mathcal{L}_n)=(-1)^{n}\mathcal{L}'_n  $, where $\mathcal{L}'_n$ is obtained from $\mathcal{L}_n$ by changing $(\alpha,\rho,\gamma)$ into  $(\alpha^{-1},-\rho,-\gamma)$.   Thus, if $W_k(\alpha,\rho,\gamma)$ denotes the Whittaker vector at level $k$ computed using $\kappa_n=\sqrt{2}$, then that computed using $\kappa_n =(-1)^{n-1}\sqrt{2}/\alpha$ is equal to $(-1)^kW_k(\alpha^{-1},-\rho,-\gamma)$.  }

In the expression for the norm, the value of $\beta$ at which the scalar product of the Jack polynomials is evaluated (that is, the value of $\beta$ in $\LL P_\la^{(\a)}| P_\mu^{(\a)}\RR_\beta$) depends crucially upon the coefficient relating $a_{-n}$ to $p_n$. With \eqref{defo}
one finds that $\beta=-{\kappa^2} \a$.
The argument goes as follows: using the relations \eqref{defo}, we get
\beq
\begin{split}
&{\mathcal{ L}}_{1}(\rho)= {\kappa}\sqrt{\a}(\rho-\gamma) \d_1 -\sum_{n>0}(n+1)\, p_{n}\, \d_{n+1}  ,  \\
&{\mathcal{ L}}_{-1}(\rho)= \frac1{{\kappa} \sqrt{\a}} (\rho+\gamma)  p_1- \sum_{n>0}n\, p_{n+1}\, \d_{n} .
\end{split} \eeq
Now enforce
\beq \LL \, \mathcal{L}_{1}(-\rho)f\,\big| \, g\, \big\RR_\beta = \LL \, f\,\big| \, \mathcal{L}_{-1}(\rho)g\, \big\RR_\beta
,\eeq 
using
$f=p_1^\ell$ and $g=p_1^{\ell+1}$. This yields $ \beta/\a=-{\kappa^2}$ as claimed. Therefore, it is possible to get $\beta=-\a$ also in the Virasoro case, but we then lose the possibility of obtaining an explicit expression of the Jack expansion coefficients. 
%
Clearly, it would be interesting to find a good argument that would fix {\it a priori} the relationship between the free-field modes and the power sums that yields nice factorized coefficients in the (s)Jack basis.

Finally, this discussion shows that it is a noteworthy property of the supersymmetric case that the corresponding value of $\kappa$ for which the coefficients  of the degenerate Whittaker vector factorize is 1 instead of 2, allowing the derivation of an explicit expression for its norm  at $\a=1$.
\sap

\section{Relation to the instanton formula}
The proper way of modifying the original AGT conjecture in order  to  link the four-dimensional supersymmetric $SU(2)$ gauge theory to superconformal blocks turns out to restrict the pure-gauge instanton partition function to the $\Z_2$-symmetric sector  \cite{BF} (see also \cite{BBB,BMTa,Ito,coset}).  The resulting  expression for the  instanton partition function reads:
\beq
Z(\textsf{a};q)=
\sum_{k\in \N/2}q^{k} \sum_{\substack{\vec Y\\N_+=k+\e\\N_-=k-\e}} \frac1{Z^{\text{sym}}_{\text{vec}}(\vec{\textsf{a}},\vec Y)}\qquad\qquad (\e=0,\tfrac12).
\eeq
It is expressed in terms of the data of a pair of Young diagrams $\vec Y=(Y_1,Y_2)$, each drawn on a chessboard with top-left box white, with total number of boxes $|Y_1|+|Y_2|=N_++N_-$, where $N_+$ (resp. $N_-$) is the number of white (resp. black) boxes. Terms with integer (resp. half-integer) powers of $q$ have $N_+-N_-=0 \;({\rm resp.~} N_+-N_-=1 )$, in which case we take $\e=0$ (resp. $\e=\tfrac12$). The quantity $Z^{\text{sym}}_{\text{vec}}$ is given by
$$Z^{\text{sym}}_{\text{vec}}(\vec{\textsf{a}},\vec Y)=\prod_{\a,\beta=1}^2\prod_{s\in {}^\diamond Y_\a(\beta)}E(\textsf{a}_\a-\textsf{a}_\beta,Y_\a,Y_\beta|s)\big(Q-E(\textsf{a}_\a-\textsf{a}_\beta,Y_\a,Y_\beta|s)\big)$$
where $Q=b+b^{-1}$, $\textsf{a}_1=-\textsf{a}_2=\textsf{a}$ 
and
$$E(\textsf{a}_\a-\textsf{a}_\beta,Y_\a,Y_\beta|s)=\big (\textsf{a}_\a-\textsf{a}_\beta+b(l_{Y_\a}(s)+1)-b^{-1}a_{Y_\beta}(s)\big )$$
with $l_Y(s)$ et $a_Y(s)$ being respectively the leg and the arm of the box $s\in Y$.
The set ${}^\diamond Y_\a(\beta)$ is defined as follows
$${}^\diamond Y_\a(\beta)=\{s\, |\, l_{Y_\a}(s)\ne a_{Y_\beta}(s)\;\text{mod}\, 2\}.$$
%
The SCFT version of the AGT conjecture is thus \cite{BF} (the factor $2^{-2k}$ is absent there):
\beq Z({\tt a};q)=\sum_{k\in\N/2} q^k 2^{-2k} \big( \,|k \rangle \, , \, |k \rangle \, \big)_{c,h}\,,
\eeq
or, equivalently, at a fixed order,
\beq 
\sum_{\substack{\vec Y\\N_+=k+\e\\N_-=k-\e}}\frac1{Z^{\text{sym}}_{\text{vec}}(\vec{\textsf{a}},\vec Y)}= 2^{-2k}\big( \,|k\rangle \, , \, | k\rangle \, \big)_{c,h}. \eeq
We have verified this relation  up to 12 instantons.

Let us specialize this expression for $c=3/2$, so that $b=i$ ($Q=0$) and $\rho = i\textsf{a}$. In this case, we
end up with the identity
\beq 
\sum_{\substack{\vec Y\\N_+=k+\e\\N_-=k-\e}}  \frac1{Z^{\text{sym}}_{\text{vec}}(\vec{\textsf{a}},\vec Y)\big|_{b=i}}= \frac{1}{ 2^{2k}}\,{ \sum_{\substack{\La\\\textrm{level}(\La)=k}}  }
w_\La(1,- i\textsf{a})\, w_{\La'}(1,  i\textsf{a})\,(-1)^{|\Lambda^\circledast|}.
\eeq
The rhs provides an alternative closed-form expression for the $\Z_2$-symmetric instanton partition function (for $b=i$) in terms of a sum over super-diagram's data. We have not been able to prove this identity but given its highly nontrivial character, we expect its presentation to be of interest.

\end{document}